\documentclass[aps,prb,amsfonts,reprint,showpacs,floatfix,unsortedaddress,superscriptaddress]{revtex4-1}

\usepackage[english]{babel}
\usepackage{graphicx}
\usepackage{dcolumn}
\usepackage{bm}
\usepackage{xfrac}

\begin{document}

\title{Dirac cones in Two-dimensional Borane}

\date{\today}
\author{Miguel Martinez-Canales}
\affiliation{SUPA, School of Physics and Astronomy \& EPCC, University of Edinburgh, Peter Guthrie Tait Road, Edinburgh EH9 3FD, United Kingdom}
\affiliation{Department of Physics and Astronomy, University College London, Gower Street, London WC1E 6BT, United Kingdom}
\email{miguel.martinez@ed.ac.uk}

\author{Timur R.\ Galeev}
\affiliation{Department of Chemistry and Biochemistry, Utah State
University, Old Main Hill 0300, Logan, UT 84322-0300, USA}

\author{Alexander I.\ Boldyrev}
\affiliation{Department of Chemistry and Biochemistry, Utah State
University, Old Main Hill 0300, Logan, UT 84322-0300, USA}

\author{Chris J. Pickard} 
\affiliation{Department of Physics and Astronomy, University College London, Gower Street, London WC1E 6BT, United Kingdom}
\affiliation{London Institute for Mathematical Sciences, 35a South St., Mayfair, London W1K 2XF, United Kingdom}
\affiliation{Department of Materials Science and Metallurgy, University of Cambridge, 27 Charles Babbage Road, Cambridge CB3 0FS, United Kingdom}
\email{cjp20@cam.ac.uk}

\begin{abstract}
We introduce two-dimensional borane, a single-layered material of BH 
stoichiometry, with promising electronic properties.
We show that, according to Density Functional Theory calculations, 
two-dimensional borane is semimetallic, with two symmetry-related Dirac cones 
meeting right at the Fermi energy $E_f$.
The curvature of the cones is lower than in graphene, thus closer to the 
ideal linear dispersion.
Its structure, formed by a puckered trigonal boron network with hydrogen atoms
connected to each boron atom, can be understood as distorted, hydrogenated
borophene (Science \textbf{350}, 1513 (2015)). 
Chemical bonding analysis reveals the boron layer in the network being bound by
delocalized four-center two-electron ${\sigma}$ bonds. 
Finally, we suggest high-pressure could be a feasible route to synthesise
two-dimensional borane.
\end{abstract}

\pacs{73.22.-f, 71.15.Mb, 73.61.-r}

\maketitle

%%%%%%%%%%%%%%%%%%%%%%%%%%%%%%%%%%%%%
%%%%%%%%%%                 %%%%%%%%%%
%%%%%%%%%%  Introduction   %%%%%%%%%%
%%%%%%%%%%                 %%%%%%%%%%
%%%%%%%%%%%%%%%%%%%%%%%%%%%%%%%%%%%%%

The discovery of graphene \cite{Novoselov2004,Novoselov2005}, the thinnest, 
one-atom-thick, planar carbon material, has fueled interest in other reduced
dimensionality systems which may have emerging properties. 
Carbon's close neighbour, boron, is also known to form honeycomb layers, as 
part of the graphite-like structure of the high-temperature superconductor MgB$_2$
\cite{Nagamatsu2001}. Here, boron atoms acquire an electronic configuration
similar to that of carbon thanks to the electrons donated by  interlayer 
magnesium atoms.
In fact, boron showcases a rich variety of bonding, from covalent to delocalized
multicentre bonds. 

Theoretical interest in all-boron 2D structures has existed for a while
\cite{Boustani1999,Evans2005,Tang2007,Tang2009,Yang2008,Zope2011,Penev2012,
SwarmOpt2012}, to cite but a few.
The hexagonal ${\alpha}$--sheet first proposed by Tang and Ismail-Beigi \cite{Tang2007} 
was thought to be the most stable 2D allotrope of boron. 
However, recent work has challenged this: first, Wu \emph{et al.} showed that the 
$\alpha$--layer was dynamically unstable \cite{Wu2012}. Then, Zhou and 
collaborators found a low-dimensionality allotrope of considerably lower energy
\cite{Zhou2014}. The synthesis of atomically thin boron allotropes has proven
challenging. But eventually, Mannix and coworkers \cite{Mannix1513} reported
last year the growth of borophene, a 2D allotrope of boron, on a silver substrate.
Against predictions \cite{Zhou2014}, borophene was found to be a highly 
anisotropic metal. Very recently, Feng and collaborators reported experimental
evidence of Dirac fermions in borophene \cite{Feng2017}.

Much effort has been devoted to design and manufacture materials with 
graphene-like properties,
especially on group IV systems such as graphyne \cite{Malko2012} and
silicene \cite{Lebegue2009, Vogt2012}. 
Bi$_{1-x}$Sb$_x$ thin-films have also seen active research \cite{Tang2012}.
%
% One example is graphyne \cite{Malko2012}, and significant body of work exists
% already for silicene \cite{Lebegue2009}, including experimental evidence of its
% formation on the Ag(111) surface \cite{Vogt2012}. Bi$_{1-x}$Sb$_x$ thin-films 
% are another promising route, in this case without group IV elements
% \cite{Tang2012}.
Attempts to control the electric properties of such layers has led to research
hydrogenated layers, such as graphane  \cite{Sofo2007}. Graphane is a fully
saturated hydrocarbon, with carbon forming an $sp^3$ bonded buckled honeycomb
network. Elias \textit{et al}.\ have reported the synthesis of graphane by 
reversible hydrogenation of graphene \cite{Elias2009}, but the efficiency of 
this method is contested \cite{Poh2012}.

One may wonder what properties a hypothetical hydrogenated boron layer will
exhibit. It is not unreasonable to expect such layer to be semimetallic or
insulating, as two electrons might form a B-H $\sigma$ bond. 
In order to answer this question, we have performed structure search on 
volume-restricted boron hydride, in order to focus on low-dimensionality
systems.
We present an analysis of a particularly stable layered two dimensional BH 
structure, which we have identified from a structure search consisting of 
many thousands of individual samples.
Compound searches were restricted to $(BH)_x$ stoichiometry, because this 
results in high hydrogenation while allowing for a fully-bonded boron network.
%Additionally, boron hydrides are chemically rich systems, which may result in
%easier synthesis pathways, or in promising energetics.

Based on our searches, we introduce a two-dimensional borane phase, with the 
formula BH, which we shall henceforth call 2D borane. 
We have computed the structural, electronic and vibrational characteristics of 
this material, characterised by the crystal structure in Table \ref{Table1}. 
%%% Referee A: remove mentions of Hydrogen storage %%%%%%
% 2D borane could find potential applicability as a hydrogen storage material owing 
% to its high hydrogen storage capacity, following from the light atomic mass of 
% boron. 
% Stacking layers of 2D borane results in a crystal of symmetry \textit{Ibam} 
% with a volumetric hydrogen capacity of 0.139 kg H$_2$/l and a gravimetric 
% capacity of 8.5 wt. \% H.
% Both values are higher than the DOE goals of volumetric and 
% gravimetric densities of 7.5 wt \% H and 0.070kg hydrogen/l, respectively 
% \cite{DOE}.

\begin{table}[tbh]
  \begin{tabular}{c c c}
  \hline
  \hline
\multicolumn{2}{c}{ $a =10.00$ \AA, $b = 3.3662$ \AA, } &
$\alpha = \beta = \gamma = 90.0^{\circ}$ \\
\multicolumn{2}{c}{ $c = 3.0298$ \AA} & \\
\hline
 Atom & Orbit & Fractional Coordinates\\
 B &  $4d$ &  (0.4554 \ 0.4027 \ 0.25)\\
 H &  $4d$ &  (0.6599 \ 0.5110 \ 0.75)\\
\hline
\hline
\end{tabular}
\caption{Structural parameters of 2D borane (\emph{Pbcm} symmetry). The long 
    edge has been chosen to minimize layer interactions. The vacuum
    along X axis is imposed by the standard crystallographic setting.
    The layered \textit{Ibam} phase is recovered by adding a centering at 
    $(\sfrac{1}{2}, \sfrac{1}{2}, \sfrac{1}{2})$, and $a = 10.669$ \AA.
    \label{Table1}}
\end{table}

%%%%%%%%%%%%%%%%%%%%%%%%%%%%%%%%%%%%%
%%%%%%%%%%                 %%%%%%%%%%
%%%%%%%%%%  Methods        %%%%%%%%%%
%%%%%%%%%%                 %%%%%%%%%%
%%%%%%%%%%%%%%%%%%%%%%%%%%%%%%%%%%%%%

The search for low-enthalpy (BH)$_x$ stoichiometry 2D structures was performed 
using \textit{ab initio} random structure searching (AIRSS) 
\cite{airss1,airss2} at the density functional theory (DFT) level. 
This technique has been successfully applied to a series of H-based systems, 
such as hydrogen \cite{Pickard2007,Pickard2012}, silane \cite{airss1}, water ice
\cite{Pickard_ice}, and ammonia monohydrate \cite{Fortes2009}.
We performed the searches with the CASTEP \cite{castep} plane-wave basis set 
DFT code and the Perdew-Burke-Ernzerhof (PBE) \cite{Perdew1996} Generalized 
Gradient Approximation (GGA) density functional and ultrasoft pseudopotentials
\cite{Vanderbilt1990}. 
We complemented the search results with structures taken from the Inorganic
Crystal Structure Database (ICSSD) \cite{icsd}.
The stability of the best candidate structures has been analysed by computing
the phonon dispersion spectra using density functional perturbation theory (DFPT) 
\cite{Baroni2001} as implemented in \textsc{quantum-Espresso} \cite{pwscf}. 
This code was also used for the Berry phase calculations and the symmetry
analysis. These calculations used frozen-core PAW potentials \cite{Blochl1994}

Additional technical details concerning the structure searches and other calculations
can be found in the supplementary material \cite{EPAPS}, and the relevant calculation
data is accessible in Ref.~\onlinecite{UoE-opendata-doi}.

Solid State Adaptive Natural Density Partitioning (SSAdNDP) \cite{Galeev2013} 
was used to analyze chemical bonding. SSAdNDP is a method to interpret bonding 
in periodic lattices in chemically intuitive terms such as Lewis-type lone pairs
and two-center bonds, as well as multi-center delocalized bonds. It is an 
extension of the AdNDP algorithm \cite{Zubarev2008} to periodic systems and as 
such was derived from a recently introduced periodic implementation 
\cite{Dunnington2012} of the Natural Bond Orbital (NBO) analysis
\cite{Foster1980, Reed1985, Reed1988, Weinhold2005}. A more detailed description
of the algorithm may be found elsewhere \cite{Galeev2013}. 
A plane-wave DFT calculation was performed using VASP 4.6 \cite{Kresse96-2},
using the PAW PBE pseudopotentials \cite{Kresse99}, summing over a $3 \times 11 \times 11$ 
k-point Monkhorst-Pack grid \cite{Monkhorst76}. 
A projection algorithm \cite{Dunnington2012} was used to obtain the 
representation of the PW DFT results in the 6-31G(dp) AO basis set prior to 
performing the SSAdNDP analysis. 
VESTA \cite{Momma11} was used for visualizations.

\begin{figure*}[t]
  \centering
  \includegraphics[height=4.6cm]{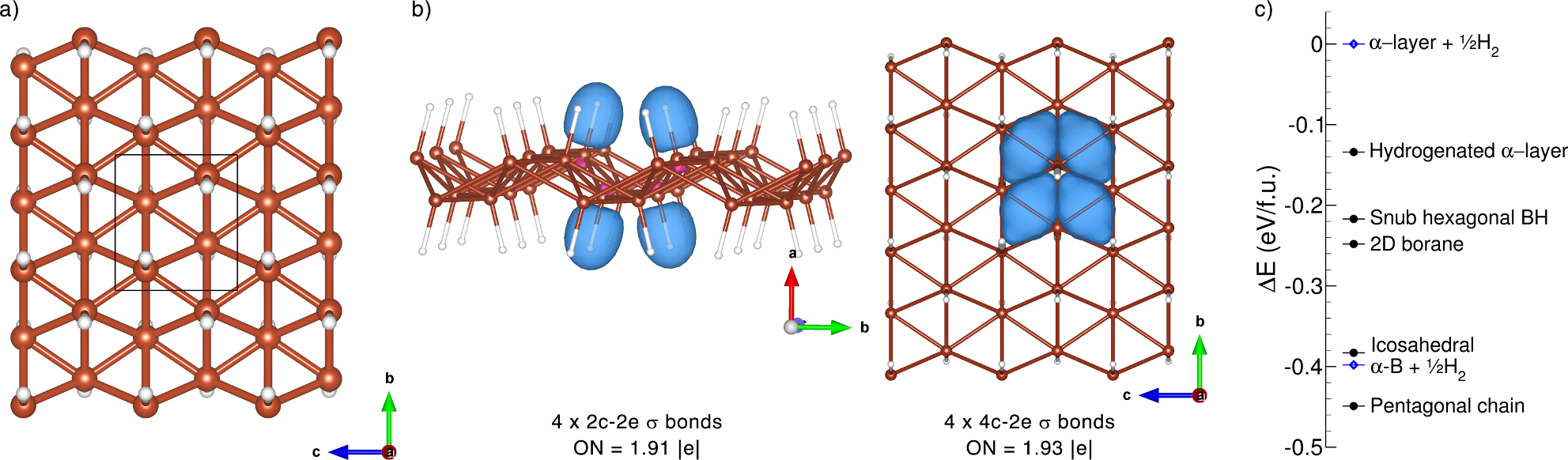}
  \caption{(color online) \textit{(a)} Structure of the proposed 2D borane. The
    boron atoms are in red and the hydrogen atoms are in gray. The unit cell is
    shown in black.
    \textit{(b)} SSAdNDP chemical bonding pattern in 2D borane: four 2c-2e B-H 
    ${\sigma}$ bonds and four 4c-2e ${\sigma}$ revealed per unit cell.
    \textit{(c)} Energies of selected borane phases, using ${\alpha}$-layer + 
    $\sfrac{1}{2}$H$_2$ as the reference. 
    \label{fig:struct}}
\end{figure*}

%%%%%%%%%%%%%%%%%%%%%%%%%%%%%%%%%%%%%
%%%%%%%%%%                 %%%%%%%%%%
%%%%%%%%%%  Results        %%%%%%%%%%
%%%%%%%%%%                 %%%%%%%%%%
%%%%%%%%%%%%%%%%%%%%%%%%%%%%%%%%%%%%%

A particularly stable layered structure found for the BH stoichiometry consisted 
of weakly interacting 2D borane layers. Each layer can be seen as a puckered 
trigonal boron network with a hydrogen atom connected to each boron atom. 
The H atoms are located on both sides of the boron layer, and arranged in 
an alternating zigzag fashion. The multilayer system has \emph{Ibam} symmetry,
while 2D borane itself belongs to the \emph{Pbam} tile group. 2D borane is 
displayed in \mbox{Fig.\ \ref{fig:struct}}. 
By removing the other 2D borane layer and relaxing the structure along the 
plane, we have checked the single layer system is mechanically stable, while 
its energetics change by less than 1 meV/atom. It should be noted that the 
actual boron network is a slight distortion from the already-synthesised 
borophene \cite{Mannix1513}. This increases our confidence in the feasibility
of this material.

The B--B lengths are 1.83--1.91 \AA, appreciably longer than B--B bond predicted 
for the ${\alpha}$--sheet (1.67 \AA), but in line with those of $Pmmm$ boron 
\cite{Zhou2014}. This is due to the fact that fewer
electrons participate in the bonding between boron atoms (one electron per
boron atom is now used for the formation of the B--H bond).
The B--H bonds are 1.19 \AA, which is close to the average B--H distance found 
in boron hydrides and their derivatives \cite{Lipscomb1963}. Hydrogenation of 
graphene to graphane leads to distortion of the carbon network from planarity, 
but the connectivity between atoms in the hexagons is preserved. The network of
boron atoms in 2D borane is however completely rearranged compared to that of 
either the ${\alpha}$--sheet, $Pmmm$ or $Pmmn$ boron \cite{Zhou2014}. 
Like the latter, 2D borane has no regular vacancy pattern, a feature that is
thought to be key to stabilise 2D boron networks \cite{Penev2012}. 
We emphasise, again, that the puckered hexagon arrangement is very similar
to that of borophene \cite{Mannix1513}.

The goal of the SSAdNDP analysis is to obtain a bonding pattern with the most 
localized bonds having occupancies close to 2 electrons and consistent with the
symmetry of the system. Thus, the search is first performed for lone pairs (1 
center – 2 electron, 1c-2e, bonds), followed by 2c-2e, 3c-2e\ldots\ $n$c-2e bonds until
the number of revealed bonds equals the number of electron pairs per unit cell. 
SSAdNDP search revealed no lone pairs in 2D borane. As expected, covalent 
$2c-2e\ {\sigma}$ bonds between hydrogen and boron atoms were found with occupation 
numbers (ON) 1.91 $|e|$, close to the ideal 2.00 $|e|$. The search revealed
no 3c-2e bonds. Instead, 4c-2e ${\sigma}$ bonds connecting boron atoms were 
found with ON 1.93 $|e|$. Thus, the four 2c-2e and four 4c-2e bonds revealed per 
unit cell account for all 8 electron pairs and no other bonding elements can be 
found in 2D borane.
Our analysis does not support the presence of $\pi$ bonds in the system.
The results of the SSAdNDP chemical bonding analysis are 
shown in Figure 1b. 
% Interestingly, the bonding in the all-boron 
% ${\alpha}$--sheet consists of in-plane 3c-2e and 4c-2e ${\sigma}$ bonds in the
% filled hexagons and out-of-plane $\pi$ bonds, \cite{Galeev2011, Galeev2013} 
% hence rationalizing the presence and arrangement of vacant hexagons in its 
% structure. In the case of the proposed 2D borane, no ${\pi}$ bonds are present: 
% out-of-plane ${\sigma}$ B-H bonds are formed instead, leaving two valence 
% electrons on each boron atom and leading to a rearranged boron network with 
% 'in-plane' 4c-2e ${\sigma}$ bonds.

\begin{figure}
  \centering
  \includegraphics[width=0.44\textwidth]{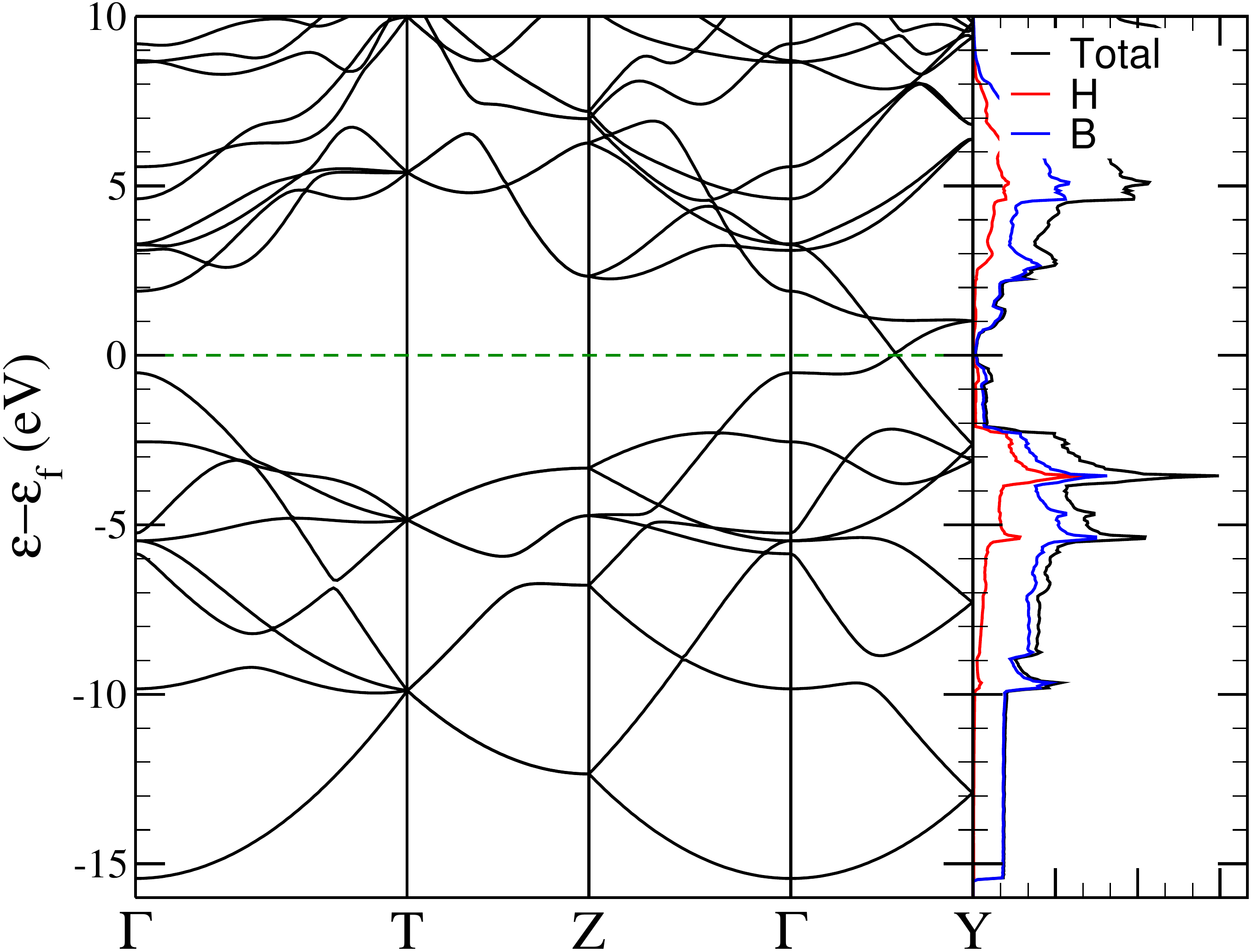}
  \caption{(color online) Band structure and density of states (DOS) of
  the single-layer 2D borane. The DOS has been computed using OptaDOS
	\cite{optados}} \label{fig:bands}
\end{figure}

We have calculated and analyzed the band structure and density of states (DOS)
of 2D borane, shown in \mbox{Fig.\ \ref{fig:bands}}. Shown in \mbox{Fig.\ 
\ref{fig:bands}}, the valence and conduction bands approach smooth-sided cones 
meeting at the Fermi energy $E_f$. 
We ruled out the possibility of additional symmetry-inequivalent cones, or a 
more complex band architecture, by computing the Fermi line of 2D borane 
\cite{EPAPS}.
Additionally, a plot of energy isosurfaces of the valence and conduction bands
shows two equivalent mirrored elliptic cones near $E_f$ \cite{EPAPS}. There is
no other state near $E_f$. The conduction and valence bands only meet at 
$\mathbf{k}_d =  \pm 0.2885 \mathbf{b}_2$,
where $\mathbf{b}_i$ represent the reciprocal lattice vectors.
In contrast, the all-boron ${\alpha}$--sheet is clearly metallic \cite{Yang2008}.

The band structure of 2D borane is related to that of 6,6,12-graphyne and $Pmmn$
boron atom derivatives \cite{Malko2012,Zhou2014}:
a cone is located not at the zone boundary, but along a high-symmetry line. The 
symmetry of the point ($m2$ or $mm$, compared to $6m$ in graphene) can result in
directional the electrical properties.
The slopes of the cone along $\Gamma\mathrm{Y}$ are $+25$ and $-47$ eV$\cdot$\AA,
and the second derivatives are $21$ and $42$ eV$\cdot$\AA$^2$ respectively.
Perpendicularly from $\Gamma\mathrm{Y}$, the slope and second derivative are $\pm 52$
eV$\cdot$\AA\ and 36 eV$\cdot$\AA$^2$ respectively.
While the slopes are similar to those of graphene (34 eV$\cdot$\AA), the curvatures 
are about 4 times smaller. These curvatures are also over an order of magnitude 
smaller than those of graphynes.
These values are very promising, and so could tempt one to say that the DFT band
structure of 2D borane near $\mathbf{k}_d$ is a more ideal cone than graphene 
itself.

It is important to understand the nature of the Dirac cone in 2D borane. In 
contrast to graphene, the degeneracy at $E_f$ is not entirely dictated by 
symmetry; it is accidental in the group theory sense. 
However, it is possible to build an effective Hamiltonian around $E_f$ and 
$\mathbf{k}_d$. From the symmetry of $Pbcm$ borane and the position of 
$\mathbf{k}_d$, we have a rectangular lattice with mirror plane and 2-axis
symmetry at $\mathbf{k}_d$. Van Miert and Smith showed that 
\cite{VanMiert2016}, in such cases, the effective Hamiltonian
\[
  H = \sum_k g(k) (s_k^\dagger s_k - p_k^\dagger p_k) + \sum_k h(k) s_k^\dagger p_k + \textrm{H.c.},
\]
can be applied to model Dirac cone physics near $E_f$, with 
$g(k_y,k_z)=\epsilon_0 + V_{nn,y} \cos k_y + V_{nn,z}\cos k_z$ and 
$h(k_y,k_z) = iV_{sp} \sin(k_z)$. We note that such effective hamiltonian 
must be built with bands of strong Boron character, disentangling the effects 
of Hydrogen-like hybridization. A projection of the character of the bands on 
atomic $s$ and $p$ orbitals is shown in the supplementary material. 
Building an effective tight-binding hamiltionan that accurately reproduces the borane 
bands is complicated by coupling between the BH $\sigma$--like valence band 
and the Boron $p_x$-like band. 
Therefore, a realistic Hamiltonian would require resorting to, at least, a 
3--band model. The chirality of $\mathbf{k}_d$ is also
displayed in a nontrivial Berry phase of $\pi$ on a path that encloses only 
one cone. Again, since this effective Hamiltonian describes the only states near
$E_f$, transport properties will be dictated by the cone.

We will now show the cone to be resilient. Far from $\mathbf{k}_d$, the
closest state to $E_f$ lies at $\Gamma$, about 0.5 eV below. 
There must be 8 full bands below $E_f$, and symmetry imposes 
$E(\mathbf{k}_d) = E(-\mathbf{k}_d) $. Thus, only perturbations able to shift
$E(\Gamma)$ by over 0.5 eV are going to move the cone away from $E_f$. 
We tested this with symmetry-preserving H displacements: the cone remains
at $E_f$ for displacements as large as $\pm 0.1$ \AA\ (8\% of the bond length).
Moreover, isostructural BF displays a qualitatively identical band structure, 
with a single symmetry-independent Dirac cone at $E_f$. So does B$_2$HF, as 
long as the mirror plane symmetry is conserved \cite{EPAPS}.
Analysing the symmetry of bands as shown by Bradlyn \emph{et al.}\cite{Bradlyn2017} 
shows the system must be, at the very least, a semimetal: the doubly degenerate 
$\mathrm{Y}_4$ state subduces into two bands of character $A_2$ and $B_2$. At $\Gamma$,
the closest levels in energy are $\Gamma_4$ (5.5 eV below $E_f$) and $\Gamma_{11}$
(3.2 eV above $E_f$). $\Gamma_4$ is the only symmetry-compatible state in the first
24 Kohn-Sham states.  
Band structure plots from molecular dynamics snapshot positions also show the 
cone persists, except for a small anticrossing gap of about $\sim k_BT$ caused 
by the instantaneous loss of symmetry.

%%%%%%%%%%%%%%%%%%%%%
%%%%%%%%%%%%%%%%%%%%%
%%%%%%%%%%%%%%%%%%%%%

\begin{figure}[tbh]
  \centering
  \includegraphics[width=0.48\textwidth]{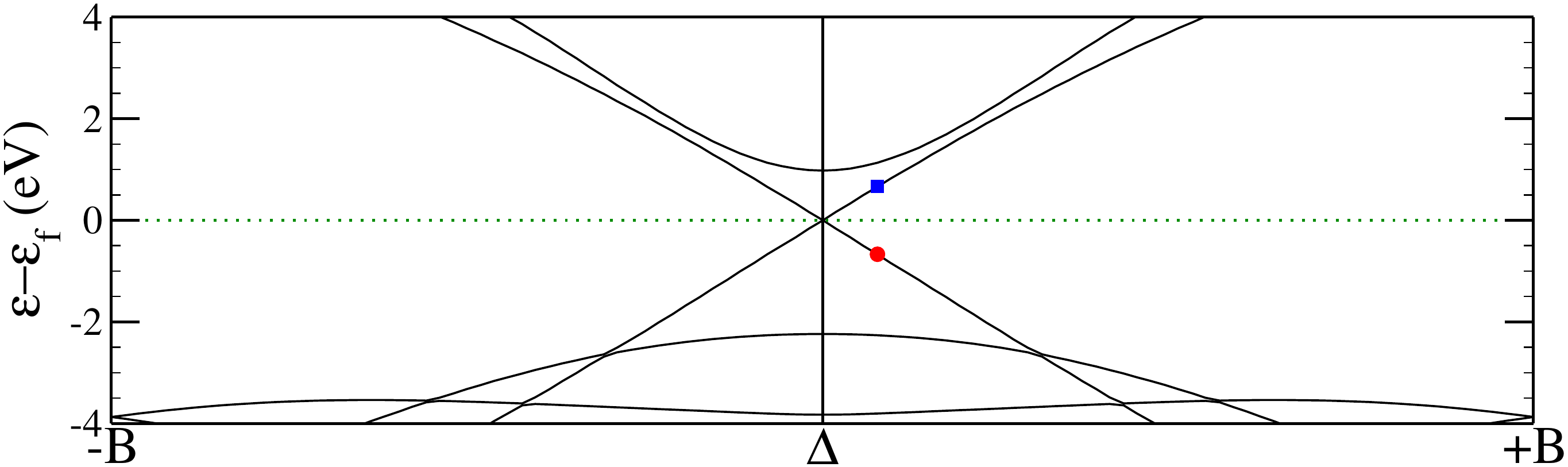}
  \includegraphics[width=0.48\textwidth]{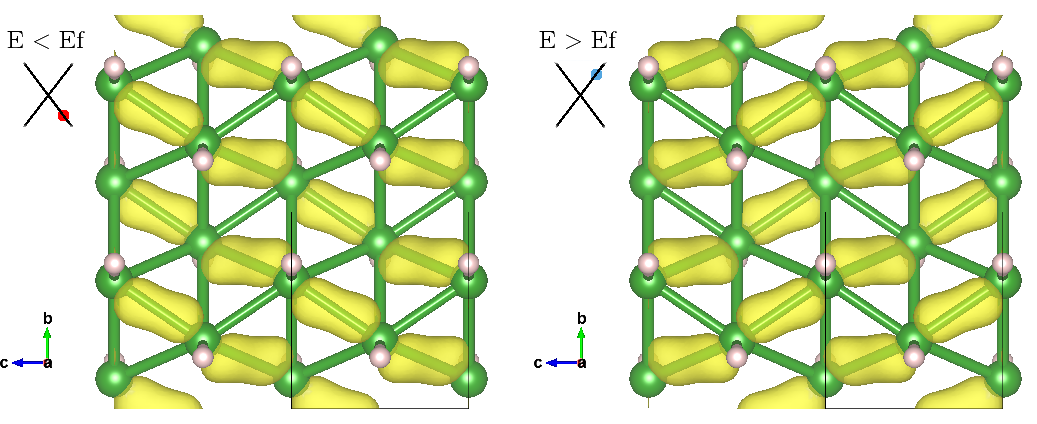}
  \caption{(color online) \emph{(top)} Detail of the bandstructure of 2D borane,
  passing through $\mathbf{k_d}$ and perpendicular to $\Gamma Y$.
  \emph{(bottom)} Charge densities associated with the highlighted k points in the
  band structure.\label{fig:detail}}
\end{figure}

In order to better understand the states that may form this effective 
Hamiltonian, we have probed the states in bands 8 and 9 for  $\mathbf{k} = 
\mathbf{k}_d + \delta\mathbf{k}$ for small $|\delta\mathbf{k}|$. 
The HOMO and LUMO states when $\delta\mathbf{k} \parallel \mathbf{b}_3$ 
are two mirrored zig-zag chains along the diagonal B-B directions. 
As $\delta\mathbf{k}$ rotates around the cone, the HOMO and LUMO charge 
densities are consistent with a linear combination of the original zig-zag 
states. 
Figure \ref{fig:detail} shows a detail of the bandstructure near $\mathbf{k}_d$,
as well as the respective HOMO and LUMO charge densities.
In an effective Dirac Hamiltonian, the hopping operator would transfer electrons
from one set of zig-zag chain to the mirrored one.
One important difference with symmetry-related crossings is that the pseudospin 
states must have different symmetry characters. If the characters were the same,
the resultant band anticrossing would turn the system into a small gap 
semiconductor.
One peculiar feature of 2D borane is the mobility of the cone: $\mathbf{k}_d$
is allowed to move along $\Gamma Y$, and does so when compressed or strained. 
This opens up the possibility of tuning the position of $\mathbf{k}_d$.

We have tested the dynamical stability of 2D borane, and found no unstable 
phonon modes exist in the Brillouin zone. Additionally, the lattice dynamics
of the multilayered \emph{Ibam} borane have also been confirmed to be stable.
The relevant phonon dispersion curves and the details of the calculations can
be found in the supplementary material \cite{EPAPS}. 
2D borane displays a quadratic acoustic mode responsible of the layer ripples 
in the long wavelength limit.
As in graphene, ripples should appear on 2D borane. The structure of the layer
is not strictly 2-dimensional, and ripples may not play a pivotal role on the
stability of 2D borane.
Finite temperature and anharmonic effects do not destabilise 2D borane: we have
performed molecular dynamics simulations for 3 ps at 300 K, in 6x6 supercells 
and with 1 fs timestep \cite{EPAPS}. 2D borane remained stable throughout the
simulation.
Finally, the phonon spectra of 2D borane and \emph{Ibam} borane are essentially 
identical. This indicates a very weak interaction between layers. Should the 
\emph{Ibam} borane be synthesized, exfoliation will be a viable mechanism to
obtain 2D borane.

The energetics of 2D borane and other BH compounds have also been analyzed, and
the results are shown in Fig.\ \ref{fig:struct}c.
Of all the various layered BH compounds analysed, 2D borane had the lowest 
enthalpy. Overall, the lowest energy structure we found in our searches is a 
ribbon of buckled pentagonal tiles, about 200 meV/BH more stable than our 
layered phase. 
The best molecular structure was a cubic arrangement of B$_{12}$H$_{12}$
icosahedra. This system is very charge-deficient and as such has only been 
observed in nature as a doubly charged anion, either as a part of a salt 
or in solution 
\cite{Pitochelli1960}. The crystal structures of these systems can be found on 
the supplementary information \cite{EPAPS}.

The large hydrogen content of 2D borane makes it a candidate for hydrogen 
storage, and so we compared the energetics of borane with the segregated phases.
As seen in Fig.\ \ref{fig:struct}, 2D borane, as well as some other layerings, 
are more stable than ${\alpha}$--layer B $+\sfrac{1}{2}\mathrm{H}_2$.
In order to find the best candidate for the hydrogenated $\alpha$--layer we
performed some additional directed searching \cite{EPAPS}. However, the best 
fully hydrogenated $\alpha$--layer we found has a higher energy than 
$\alpha$--layer $+\sfrac{1}{2}\mathrm{H}_2$, and thus higher than 2D borane too.
We also found a hydrogenated analogue of the snub boron sheet predicted by Zope 
and Baruah \cite{Zope2011}. It is, however, 32 meV/formula unit less stable than
2D borane.
We have computed the full $\mathrm{B}_x\mathrm{H}_y$ hull, including large 
experimental phases of various stoichiometries, as seen in the supplementary
information \cite{EPAPS}.
We found molecular diborane (B$_2$H$_6$) not to be a thermodynamically stable 
stoichiometry. At the PBE level, the only stable stoichiometry was 
B$_9$H$_{11}$.
We note the stable phase is formed by boron-defective icosahedra.
As expected, 2D borane would decompose, but not into diborane plus 
$\alpha$--layer. In the layered case, the reaction
\[
\textrm{2D borane} \rightarrow \alpha\textrm{--layer} + \frac{1}{2}H_2
\]
requires an energy of 0.496 eV/H$_2$. It is in the 0.2-0.6 eV range, the optimal
binding energy for an ambient conditions hydrogen storage material 
\cite{Kim2006,Sorokin2011}.

As a final remark, a letter on boron hydrides under pressure, published during 
the review process, reported the stabilization of the BH stoichiometry \cite{Hu2013}. 
Hu \emph{et al.} independently also found the multilayered $Ibam$ phase, which
they report to become the stable BH structure under pressure. This suggest 
high-pressure synthesis of $Ibam$ BH, followed by exfoliation, is a plausible
mechanism for the synthesis of 2D borane.

In conclusion, in this work we introduce 2D borane, a single-layered BH 
material. It is the most stable layered phase found in our structure searches.
The energetics show 2D borane is stable towards decomposition to $\alpha$--layer +
$\sfrac{1}{2}$H$_2$. It is formed by a puckered triangular boron network, in 
which hydrogen atoms connect to each boron site from alternating sides of the 
network, forming zigzag like chains. 
The band structure and density of states of 2D borane show two symmetry related
smooth-sided cones with the Dirac points right at the Fermi energy. The cone
structure is protected by the electron counting sum rule and is located along a 
high-symmetry line. As a result, it can move along $\Gamma Y$.
2D borane also shows remarkably small curvatures even in symmetry-unrestricted 
directions, suggesting this material may have graphene-like electronic 
properties. 
The SSAdNDP analysis revealed that the boron atoms are bonded
by delocalized 4c-2e ${\sigma}$ bonds. 
We hope that our work will motivate an experimental search for this material.

\acknowledgments

C.J.P. and M.M.C. \ acknowledge financial support from EPSRC grants EP/K013688/1
and EP/G007489/2 (UK) and the use of
the UCL Legion High Performance Computing Facility as well as HECToR and Archer, 
the UK's national high-performance computing services. 
CJP further acknowledges Department of the Navy Grant N62909-12-1-7109, issued 
by Office of Naval Research Global.
The work at USU was
supported by National Science Foundation (grant CHE-1664379). Compute, storage 
and other resources from the Division of Research Computing in the Office of 
Research and Graduate Studies at Utah State University, and the support and 
resources from the Center for High Performance Computing at the University of 
Utah are gratefully acknowledged

\bibliographystyle{apsrev4-1}
\bibliography{borane}

\end{document}